\documentclass[conference]{IEEEtran}

\IEEEoverridecommandlockouts                              

\usepackage{amsmath}
\usepackage{parskip}
\usepackage{listings}
\usepackage{indentfirst}
\usepackage{graphicx}
\usepackage{float}
\usepackage{extarrows}
\usepackage{amssymb}
\usepackage{amsthm}
\usepackage{subfigure}
\usepackage{epstopdf}
\usepackage{caption}
\usepackage{cite,color}
\usepackage{todonotes}
\usepackage{tikz,pgfplots}
\usepackage{xcolor}
\usetikzlibrary{automata,positioning}
\usetikzlibrary{arrows,shapes,chains}
\usepgflibrary{patterns}

\begin{document}
\title{Timely Broadcasting in  Erasure Networks: Age-Rate Tradeoffs}
\author{Xingran Chen,\IEEEmembership{}
Renpu Liu*, \IEEEmembership{} 
Shaochong Wang*, \IEEEmembership{} 
Shirin Saeedi Bidokhti \IEEEmembership{}
 	
\IEEEcompsocitemizethanks {\IEEEcompsocthanksitem Xingran Chen, Shaochong Wang, Renpu Liu and Shirin Saeedi Bidokhti are with the Department of Electrical and System Engineering,  University of Pennsylvania, PA, 19104.\quad 
E-mail: \{xingranc, tainerl, shaowang, saeedi\}@seas.upenn.edu
}
\thanks{*: contributed equally}
	
}

\IEEEoverridecommandlockouts
\newtheorem{lemma}{Lemma}
\newtheorem{note}{Note}
\newtheorem{property}{Property}
\newtheorem{theorem}{Theorem}
\newtheorem{definition}{Definition}
\newtheorem{corollary}{Corollary}
\newtheorem{proposition}{Proposition}
\newtheorem{remark}{Remark}
\newtheorem{assumption}{Assumption}
\newtheorem{example}{Example}

\maketitle
\thispagestyle{empty}
\pagestyle{empty}

\begin{abstract}
The interplay between timeliness and rate efficiency is investigated in packet erasure broadcast channels with feedback.  A scheduling framework is proposed in which coding actions, as opposed to users, are scheduled to attain desired tradeoffs between rate and age of information (AoI). This tradeoff is formalized by an upper bound on AoI as a function of the  target rate constraints and two lower bounds: one as a function of the communication rate and one as a function of the arrival rate.  Simulation results show that (i) coding can be beneficial in reducing AoI in the regime of moderate  arrival rates even without rate constraints and the benefit increases with the number of users, and (ii) AoI increases with both the target rate constraint  and the arrival rate when either is kept fixed,  but decreases with them when they are set to be equal.
\end{abstract}

\section{Introduction}\label{sec: Introduction}
The technology of Internet of Things (IoT) provides a vision for 
integrating intelligence into cyber-physical systems using real-time applications. Timeliness is key for such applications and  it has therefore emerged as a communication design criteria. There are, however,  tradeoffs between timeliness and rate which we aim to investigate  in  broadcast networks.

Timeliness is measured using the metric of {\it Age of Information (AoI)}, as introduced in~\cite{SKMGVRJK2011}.
AoI captures, at the receiving side, how much time has passed since the generation time of the latest received packet. In the past decade, Age of information has been extensively investigated for status update systems \cite{SKRYMG2012, SKRYMG_1_2012, CKSKAE2014, CKSKGNAE2016, MCMCAE2016, RYSK2019}.  From the aspect of scheduling, optimal transmission policies were proposed in \cite{IKEM2019,IKASEM2019, IKASEBRSEM2018, IKASEM2018, RTIKSKEM2018} to optimize the overall age in wireless networks. The reader is referred to \cite{AKNPVA2017},  \cite{Y2020age} for a survey on the topic.

Rate efficiency is often provided by channel coding schemes over multiple realizations of the network and it comes at the cost of large delays. It is, therefore, not clear a-priori what types of tradeoffs exist between rate  and timeliness. In coding theory, previous works have mainly studied point to point channels \cite{KCLH2016,RYENESJZ2017,PPATJC2017}. In point to point erasure channels,\cite{ENETRN2019} proves that when the source alphabet and  channel input alphabet have the same size, a  Last-Come First-Serve  (LCFS) policy with no coding is optimal. This is in contrast to channel coding schemes that provide rate efficiency by block coding.  Considering erasure channels with FCFS M/G/1 queues, \cite{HSTBEBGD2018}  finds an optimal  block length for channel coding to minimize the average age and average peak age. 
In the context of broadcast packet erasure channels (BPECs) with feedback, coding is shown to be beneficial for age efficiency with two users \cite{cxrAoI12019}. In related work, \cite{SFJY2020, SFJY2019} design optimal precoding schemes to minimize AoI in a MIMO broadcast channel with multiple senders and receivers under  FIFO  channels without packet management. Reference 
\cite{MCYES2019} analyzes the AoI in a multicast network with network coding.

In this work, we consider erasure networks and devise broadcast strategies  that are efficient both in AoI and rate. The inherent tradeoff can be explained as follows. 
On the one hand, a higher rate effectively corresponds to a smaller delay (both in the sense that the queues get emptied faster and in the sense that fewer uses of the network are needed in total to transmit a fixed number of information bits), hence may correspond to a smaller AoI. 
On the other hand, to achieve high rates with coding, we have to incur delay by waiting for the arrival/generation of other packets for the purpose of coding as well as prioritizing their transmission, and this leads to larger AoI. 
To shed light on the above tradeoff, we build on our previous work \cite{cxrAoI12019} and consider an erasure wireless network with $M$ users.
Motivated by the success of age-based scheduling in wireless networks, we propose a scheduling framework where we schedule various useful coding actions as opposed to scheduling the users. Within this framework, we can capture both rate efficiency as well as age efficiency. In particular, we  design deterministic policies that minimize the expected weighted sum of AoI (EAoI) under given target rate constraints.

The contributions of the work are summarized as follows: (i) We propose a novel framework of network AoI on the broadcast channels under transmission mechanism with coding (Section~\ref{sec: Basic model}). (ii) (Near-)optimal coding policies with uncoded and coded caching are proposed (Section~\ref{sec: Scheduling Coding Actions}, \cite[Section~IV]{cxrisit2021}). Two general lower bounds and an upper bound are derived on EAoI for any transmission policy (Section~\ref{sec: lower bound}, Theorem~\ref{thm: upper bound-1}). The bounds are functions of generation rates, erasure probabilities and target rate constraints. (iii) Simulation results reveal that (a) coding is beneficial, and the benefits increase with the number of users; (b) a good approximation of proposed policies is obtained based on maximum clique size of information graph; (c) the tradeoff between rate and AoI exists, which implies that the system has to sacrifice AoI to achieve a higher rate.

\subsection{Notation}
We use the notations $\mathbb{E}(\cdot)$ and $\Pr(\cdot)$ for expectation and probability, respectively. We denote scalars with lower case letters, e.g. $s$. Denote vectors as lower case letters with underline, e.g., $\underline{s}$. Random variables are denoted by capital letters, e.g. $S$. Sets are denoted by calligraphy letters, e.g. $\mathcal{S}$. We use $M$ to denote the number of users, $K$ to denote the time horizon. $[n]$ denotes the set $\{1,2,\cdots,n\}$. For two sets $\mathcal{A}$ and $\mathcal{B}$, $\mathcal{A}\subset\mathcal{B}$ represents that $\mathcal{A}$ is a subset of $\mathcal{B}$.

\section{System Model}\label{sec: Basic model}
 The system model extends that of \cite[Section~VI]{cxrAoI12019} to $M$ users. In particular, transmission occurs in a wireless network which we model by a Broadcast Packet Erasure Channel (BPEC) with $M$ users. In the beginning of time slot $k$, a packet intended for user $i$ is generated with probability $\theta_i$. Let $G_i(k)=1$ represent that a new packet is generated (for user $i$) in time slot $k$. So $\Pr\big(G_i(k)=1\big)=\theta_i$ for $k=1,2,\ldots$.  
 
Every broadcast packet is received at user $i$ with probability $1-\epsilon_i$, $0\leq\epsilon_i<1$, and lost with probability $\epsilon_i$.  Erasure events at multiple users can be  dependent in general. 
The {\it transmission delay} is assumed fixed and equal to one time slot. After each transmission, the transmitter receives ACK/NACK feedback from all receivers and can thus calculate and track the aging of information at each user. Let $d_i(k)=1$ if user $i$ decodes a packet of type $i$ in time $k$, and $d_i(k)=0$ otherwise.

If a packet is not received at its intended user, it can be cached by other user(s) that have received it. 
The cached packets can act as side information. Using the available feedback, the encoder can track the cached packets and  exploit them as  side information in the code design to form more efficient coded packets  that are simultaneously useful for multiple users. 
Such code designs have also appeared in \cite{MGLT2013, SSBMGCF2012, MHSSB2018, SCLIHWAV2019} for rate-efficiency.

 We call a packet {\it a coded} if it is formed by combining  more than one packets; otherwise we call it an {\it uncoded packet}. 
Consider a coded packet $x$. If user $i$ can instantaneously {\it decode} a packet that is intended for him upon successful delivery of $x$ (possibly using its locally cached packets), we call user $i$ a {\it destination} for packet $x$. A coded packet can be {\it fully decoded} by user $i$ if user $i$ extracts every uncoded packets combined within it upon successful delivery.

Depending on the available caching and coding capabilities, we can consider three class of policies: (i) policies that benefit from coding by caching uncoded packets,  (ii) policies that benefit from coding by caching general (potentially coded) packets, and (iii) policies that schedule different users and perform no caching/coding \cite{IKASEBRSEM2018, IKASEM2018} (time-sharing policies). We investigate the first class in Section~\ref{sec: Scheduling Coding Actions} and refer them as coding policies with uncoded caching. The second class, referred by coding policies with coded caching, is investigated in \cite[Section~IV]{cxrisit2021}. Time-sharing policies from benchmarks in our simulations in Appendix~\ref{sec: Benchmark}.
In this work, we consider coding policies with uncoded caching and linear network coding through XOR operations only. Coding over larger finite fields may impose larger decoding delay and is often practically less desirable.

 \subsection{A Virtual Network of Queues}\label{sec: A Virtual Network of Queues}
The idea of caching and coding on the fly is to cache overheard packets at the users and track them using feedback at the encoder  through a \emph{network of virtual queues}.

Let  $Q_i$ denote the queue of incoming packets for user $i$. If a packet chosen from $Q_i$ is transmitted and received by its intended user, it is removed from the queue. If it is not received by its intended user $i$, but received by some other users, then the packet will be cached in the  cache of those users (as side information) and tracked at a virtual queue at the encoder. The buffer size of each virtual queue is assumed infinite. Define $Q_{i,\mathcal{S}}$ as the virtual queue that tracks, \emph{at the encoder}, uncoded packets for user $i$ that are received {\it only} by the users in $\mathcal{S}$, where  $\mathcal{S}\subset [M]\backslash i$. Note that $Q_i$ \big($=Q_{i,\varnothing}$\big) is some sort of $Q_{i,\mathcal{S}}$. Queue $Q_{i,\mathcal{S}}$ contains two types of  packets: packets from $Q_i$ that are cached (received or decoded) by the users in $\mathcal{S}$, and/or uncoded packets combined within coded packets which are fully decoded.
The queues $Q_{i,\mathcal{S}}$ are defined so that the set of packets in them are disjoint.

\begin{figure}[t!]
\centering
\includegraphics[width=1.8in, height = 1.5in]{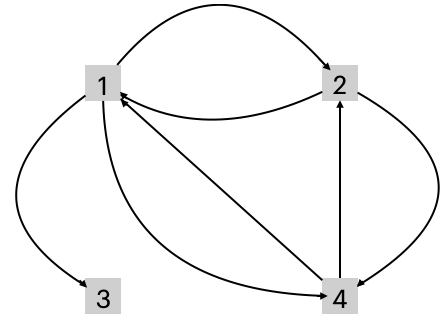}
\caption{Each clique corresponds to a coded packet.}
\label{fig:clique}
\end{figure}

Packets stored in the virtual queues at the encoder can form efficient coded packets that are simultaneously useful for multiple users. In this work, we consider linear network coding through XOR operations only. This is because for broadcast erasure channel with multiple unicast traffic, using simple coding operations leads to low decoding delay and is also practically desirable \cite{SAMGLGLT2012}. For example, consider a packet in $p_1$ in $Q_{1,\{2\}}$ and a packet $p_2$ in $Q_{1,\{2\}}$. the XOR packet $x=p_1+p_2$ is useful for both users $1$ and $2$ because user $2$ has cached packet $p_1$ and user $1$ has cached packet $p_2$ and they can therefore recover their desired packets by XORing packet $x$ with their respective cached packet. More generally, consider a set of non-empty queues $\{Q_{\tau_i,\mathcal{S}_{\tau_i}}\}_{i=1}^{\ell}$ where $\tau_i$ is a user index ($\tau_i\in [M]$) and $\mathcal{S}_{\tau_i}$ is a subset of $[M]\backslash \tau_i$. Suppose the following condition holds: 
\begin{align}
\mathcal{S}_{\tau_i}\supset \{\cup_{j=1, j\neq i}^\ell\tau_j\}\qquad \forall i=1,\ldots \ell.\label{eq:condition}
\end{align}
Then  XORing packets $$p_i\in Q_{\tau_i,\mathcal{S}_{\tau_i}}$$  forms a coded packet $x$ as follows 
\begin{align}
x=\bigoplus_{i=1}^\ell p_i
\end{align}
which is simultaneously decodable at all users $\{\tau_1,\ldots,\tau_\ell\}$.
To view condition \eqref{eq:condition} alternatively, draw a side information graph $\mathcal{G}$ with nodes $V=\{1,\ldots,M\}$. Add an edge between nodes $(i,j)$ if $Q_{i,\mathcal{S}_i}$ is non-empty for some set $\mathcal{S}_i$ that has $j$ as an element. On this graph, condition \eqref{eq:condition} corresponds to the subgraph induced by nodes $\{\tau_1,\ldots,\tau_\ell\}$ forming a clique of size $\ell$. An example with a clique of size $3$ is shown in Figure~\ref{fig:clique} with $4$ users. Consider packets $p_1\in Q_{1, \{2,3,4\}}$, $p_2\in Q_{2, \{14\}}$ and $p_4\in Q_{4,\{12\}}$. Let clique $1\leftrightarrow 2\leftrightarrow4\leftrightarrow1$ corresponds to the coded packet $x = p_1\oplus p_2\oplus p_4$.

The {\it coding actions} we consider in this section correspond to cliques on the side information graph (which has to be  updated on the fly after each transmission). In this class, maximal cliques are sufficient to consider among all cliques because sending a coded packet that corresponds to a subset of cliques is at most as useful (in terms of the users at which coded packets are decodable) as a coded packet that corresponds to a maximal clique.
Among all possible maximal cliques (the number of which can generally be on the order of $3^{\frac{M}{3}}$ \cite{JMLM1965}) we aim to choose (schedule) one that leads to a coding action with the most benefit in terms of information freshness and rate.

\subsection{Age and Rate Efficiency}\label{sec: Age and Rate Efficiency}
To capture the freshness of information, we use the metric of {\it Age of Information} (AoI) defined in \cite{AKNPVA2017}.
Denote $h_i(k)$ as  the AoI of user $i$ in time slot $k$. The age function  $h_i(k)$ increases linearly in time when no delivery for user $i$ occurs and drops with every delivery to a value that represents how old the received packet is.  If an outdated packet (for user $i$) is received (meaning that a more recently generated packet is previously received at user $i$) then the outdated packet does not offer age reduction and $h_i(k)$ keeps increasing linearly. 
\begin{definition}\label{def: AoI for queue}
	Denote the generation time of the packet received by user $i$ in time slot $k$ as $v_i(k)$. Assuming the initial state $h_i(0)=1$, the age function $h_i(k)$ evolves as follows: 
	\begin{align*}
	h_i(k)&=\left\{
	\begin{array}{ll}
	\min\{h_i(k-1)+1, k - v_i(k)\} &d_i(k)=1\\
	h_i(k-1)+1& d_i(k)=0.
	\end{array}
	\right.
	\end{align*}	
\end{definition}
The expected weighted sum of AoI (EAoI) at the users is thus given by $\mathbb{E}[J_K^\pi]$ and
\begin{align}\label{equ: BM-expected weighted sum aoi}
J_K^\pi := \frac{1}{MK}\sum_{k=1}^{K}\sum_{i=1}^{M}\alpha_i h^{\pi}_{i}(k)
\end{align}
where $\alpha_1,\alpha_2,\cdots,\alpha_M$ are weights and the superscript $\pi$ represents the communication policy.
We are interested in minimizing EAoI under some constraints on the rate of communications.
We define the {\it communication rate} to user $i$  as the number of decoded packets (intended for user $i$)  per time slot in the limit of time.
The larger the rate, the fewer packet in the network of virtual queues at the encoder.

Let $q_i$ be a strictly positive real value that represents the minimum rate requirement of node $i$. Without loss of generality, we assume that $\underline{q}=(q_1, q_2, \cdots, q_M)$ is in the capacity region. Similar to \cite{IKASEM2019}, we define the {\it long-term rate} of node $i$ when policy $\pi$ is employed as
\begin{align}\label{eq: hat q_ik}
r_i^\pi:=\lim_{K\to\infty}\frac{1}{K}\sum_{k=1}^{K}\mathbb{E}\big[d_i^\pi(k)\big].
\end{align}
Then, we express the {\it minimum rate constraint} of each individual node as
\begin{align}\label{eq: minimum throughput constraint}
r_i^\pi\geq q_i, i=1,2,\cdots, M.
\end{align}

Ultimately, we seek to schedule the coding actions in order to achieve a judicious tradeoff between the EAoI and communication rate, as outlined below. 
Combining \eqref{equ: BM-expected weighted sum aoi}, \eqref{eq: hat q_ik} and \eqref{eq: minimum throughput constraint}, the objective  is given by the following optimization problem: 
\begin{equation}\label{eq: tradeoff_objective}
\begin{aligned}
J(\underline{q}):=\min_{\pi}\quad&\lim_{K\to\infty}\mathbb{E}[J^\pi_K]\\
s.t\quad&r_i^\pi\geq q_i, i=1,2,\cdots, M.
\end{aligned}
\end{equation}

\section{Scheduling Coding Actions in Uncoded Caching}\label{sec: Scheduling Coding Actions}
In this section, we consider coding policies with uncoded caching, i.e., all  cached packets are uncoded.
We develop and analyze max-weight  policies that schedule the coding actions to optimize \eqref{eq: tradeoff_objective}. Our results can be generalized to the case where we allow the caching of coded packets as outlined in \cite[Section~IV]{cxrisit2021}.

Each coding action can be described by a set of queues, each storing multiple packets. We allow packet management in choosing which packets of the chosen queues to use to form coded packets as it reduces age without impacting rate.

In order to optimize for age, we first define the  AoI of queues $Q_{i,\mathcal{S}}$ in the virtual network of queues (at the encoder) and explain their time evolution. 
Recall that $Q_{i,\mathcal{S}}$ is the queue that contains those packets of user $i$ that are decodes only by the users in $\mathcal{S}$. Thus, if $a\in Q_{i,\mathcal{S}}$, for any $\mathcal{S}'\subset \mathcal{S}$ and $\mathcal{S}'\neq \mathcal{S}$, $a\notin Q_{i,\mathcal{S}'}$.
In addition, if $a\in Q_{i,\varnothing}$, then $a\notin Q_{i,\mathcal{S}}$ for all $\mathcal{S}\neq\varnothing$. So  the map from packets to queues is a surjection. 
From Section~\ref{sec: A Virtual Network of Queues}, the encoder decides among the following actions, denoted by $A(k)$, and defined below:
\begin{itemize}
	\item $A(k)=Q_{i, \varnothing}$: a packet is transmitted from $Q_{i, \varnothing}$;
	\item $A(k)=\oplus_{j=1}^{l}Q_{\tau_j,\mathcal{S}_{\tau_j}}$: a coded packet is transmitted that is formed by an XOR of $l$ packets, one from each of the queues $Q_{\tau_1,\mathcal{S}_{\tau_1}}, Q_{\tau_2,\mathcal{S}_{\tau_2}}, \cdots, Q_{\tau_l,\mathcal{S}_{\tau_l}}$, where $S_{\tau_l}\not\owns\tau_l$ and users $\tau_1,\tau_2,\cdots,\tau_l$ form a {\it maximal} clique on the side information graph.
\end{itemize}

\subsection{Encoder's Age of Information}\label{sec: AoI}
To capture the aging of information at the encoder, we define a notion of AoI of each virtual queue. The following Lemma is proved in Appendix~\ref{App: lem: at most 1 packet in Qii}
\begin{lemma}\label{lem: at most 1 packet in Qii}
If $p_j\in Q_{i,\mathcal{S}}$ has the generation time $k_j$, $j\in\{1,2\}$, and $k_2>k_1$, then (encoding and) transmitting $p_2$ can not be worse than (encoding and) transmitting $p_1$ in terms of AoI.
\end{lemma}

If $\mathcal{S}=\varnothing$, denote the AoI of $Q_{i, \varnothing}$ by $w_{i, \varnothing}(k)$, and the generation time of latest packet by $k'$. Based on Lemma~\ref{lem: at most 1 packet in Qii}, we define $$w_{i, \varnothing}(k) = \min\{k-k', h_i(k)\}$$ and $w_{i, \varnothing}(0)=h_i(0)$. This is to capture the fact that if $k-k'>h_i(k)$, then packets in $Q_{i, \varnothing}$ are older than the latest one recovered by user $i$, so packets in $Q_{i, \varnothing}$ are obsolete in terms of AoI in time slot $k$.
The evolution of the AoI at the queue $Q_{i,\varnothing}$ is as follows: $w_{i, \varnothing}(k)$ drops to $0$ if a new packet is generated; otherwise it increases by $1$. Thus, the recursion of $w_{i, \varnothing}(k)$ is
\begin{equation}\label{eq: recursion of w_ii}
\small
\begin{aligned}
w_{i, \varnothing}(k+1)=\left\{
\begin{aligned}
&0&& G_i(k)=1\\
&\min\{w_{i, \varnothing}(k)+1,h_i(k)+1\}&&G_i(k)=0
\end{aligned}
\right..
\end{aligned}
\end{equation} 

Before defining the AoI of $Q_{i,\mathcal{S}}$, let  $t_{i,\mathcal{S}}(k)$ be an indicator function as follows:  $t_{i,\mathcal{S}}(k)=1$  if the latest packet in $Q_{i,\mathcal{S}}$ is encoded and transmitted in time slot $k$, and is  $t_{i,\mathcal{S}}(k)=0$ otherwise. In Figure~\ref{fig:clique}, if the encoder transmits $p = a_1\oplus a_2\oplus a_4$ in time slot $k$, then $t_{1,\{2,3,4\}}(k)=1$, $t_{2,\{1,4\}}(k)=1$ and $t_{4,\{1,2\}}(k)=1$. Now we consider the AoI of $Q_{i,\mathcal{S}}$ with $\mathcal{S}\neq\varnothing$. Denote the AoI of $Q_{i,\mathcal{S}}$ as $w_{i,\mathcal{S}}(k)$. Let the generation time of the latest packet in $Q_{i,\mathcal{S}}$ be $k'$. We define $$w_{i,\mathcal{S}}(k)=\min\{k-k', h_i(k)\}$$ and $Q_{i,\mathcal{S}}(0)=h_i(0)$. Then, $w_{i,\mathcal{S}}(k)$ increases by $1$ unless $Q_{i,\mathcal{S}}$ is updated with a fresher packet. The content of $Q_{i,\mathcal{S}}$ change when packets move in other virtual queues at the encoder. For example,
If packet $a\in Q_{i,\mathcal{S}}$ is recovered by other users in $\mathcal{I}$, $\mathcal{I}\cap\mathcal{S}=\varnothing$, then $a\in Q_{i,\mathcal{I}\cup\mathcal{S}}$ and $a\notin Q_{i,\mathcal{S}}$.
Let $d_{\mathcal{S}}(k)=1$ represent that only users in $\mathcal{S}$ receive a (coded) packet. 
Let $\mathcal{S}'\subset\mathcal{S}$. The recursion of $w_{i,\mathcal{S}}(k)$ is
\begin{equation}\label{eq: w_iS}
\small
\begin{aligned}
w_{i,\mathcal{S}}(k+1) = \left\{
\begin{aligned}
&\min\{w_{i, \varnothing}(k)+1,h_i(k)+1\}&& \mathcal{P}_{i,\mathcal{S}}^{(1)}(k) \\
&\min\{w_{i,\mathcal{S}'}(k)+1,h_i(k)+1\}&& \mathcal{P}_{i,\mathcal{S}}^{(2)}(k) \\
&\min\{w_{i,\mathcal{S}}(k)+1,  h_i(k)+1\}&&\text{otherwise}
\end{aligned}
\right..
\end{aligned}
\end{equation}
where 
\begin{align*}
\mathcal{P}_{i,\mathcal{S}}^{(1)}(k) = \{d_i(k) = 0, d_{\mathcal{S}}(k)=1, t_{i, \varnothing}(k)=1\}
\end{align*}
and
\begin{align*}
\mathcal{P}_{i,\mathcal{S}}^{(2)}(k) = &\{\mathcal{S}'\subset\mathcal{S}, d_i(k) = 0, d_{\mathcal{S}\backslash \mathcal{S}'}(k)=1, t_{i,\mathcal{S}'}(k)=1\}.
\end{align*}

From \eqref{eq: recursion of w_ii} and \eqref{eq: w_iS}, the recursion of $h_i(k)$ is
\begin{equation}\label{eq: recursion of h}
\begin{aligned}
h_i(k+1)=\left\{
\begin{aligned}
&w_{i,\mathcal{S}}(k)+1& & t_{i,\mathcal{S}}(k)=1, d_i(k)=1\\
&h_i(k)+1&&\text{otherwise}
\end{aligned}\right..
\end{aligned}
\end{equation}

\subsection{Age-Rate Max-Weight Scheduling}\label{sec: Age-based Max-Weight Policy}
It is well established that coding actions can enhance the communication rate of broadcast channels \cite{MGLT2013}, and may incur additional delays.  To seek efficiency both in AoI and communication rate, similar to \cite{IKASEM2018, IKASEM2019, cxrAoI12019}, we propose Age-Rate Max-Weight (ARM) policies to minimize EAoI in \eqref{eq: tradeoff_objective} under rate constraints. 

We define the {\it age-gain} of queue $Q_{i,\mathcal{S}}$ (for user $i$), where $\mathcal{S}\subset [M]\backslash i$ as follows:
\begin{align}
\delta_{i,\mathcal{S}}(k) = h_i(k) - w_{i,\mathcal{S}}(k).\label{eq: age-gain ij}
\end{align}
The term $\delta_{i,\mathcal{S}}(k)$ quantifies how much the instantaneous user's age of information reduces upon successful delivery from the encoder's virtual queue $Q_{i,\mathcal{S}}$. If $Q_{i,\mathcal{S}}$ is empty or contains old packets, then by the definition of $w_{i,\mathcal{S}}(k)$, $\delta_{i,\mathcal{S}}(k)=0$.

Let $x_i(k)$ be the throughput debt associated with node $i$ at the beginning of slot $k$ \cite{IKASEM2019}. It evolves as follows:
\begin{align}\label{eq: xplus}
x_i(k+1) = kq_i - \sum_{\tau=1}^{k}d_i^\pi(\tau).
\end{align}
The value of $kq_i$ is the minimum average number of packets that node $i$ should have decoded by slot $k+1$ and $\sum_{\tau=0}^{k}d_i^\pi(\tau)$ is the total number of recovered packets   in the same interval. In fact, strong stability of the process $x_i^+(k)$
is sufficient to establish that the minimum rate constraint, $r_i^\pi\geq q_i$, is satisfied \cite{IKASEM2019}, \cite[Theorem~2.8]{M2010}.

Define the encoder's state in time slot $k$ as $$S(k)=\Big(\{h_i(k)\}_i, \{w_{i,\mathcal{S}}(k)\}_{i,\mathcal{S}}, \{x_i(k)\}_i\Big),$$
and the Lyapunov function $L\big(S(k)\big)$ as
\begin{equation}\label{eq: new Lyapunov}
\begin{aligned}
L\big(S(k)\big) = \sum_{i=1}^{M}\beta_i h_i(k) + \lambda\sum_{i=1}^{M}\big(x_i^+(k)\big)^2
\end{aligned}
\end{equation}
where $\beta_i, \lambda>0$. Here, the quadratic function for $x_i(k)$ is to maximize the rate \cite{IKASEM2018, IKASEM2019, IKEM2019}, and the linear function for $h_i(k)$ is to simplify the derivation. The one-slot Lyapunov Drift is defined as
\begin{equation}
\Theta(k) = \mathbb{E}\Big[L\big(S(k+1)\big)-L\big(S(k)\big)|S(k)\Big].
\end{equation}

Define  the {\it rate-gain} of user $i$ in time slot $k$ as follows:
\begin{equation}\label{eq: f_i}
\begin{aligned}
f_i(k)=\Big(\big(x_i(k)+q_i\big)^+\Big)^2-\Big(\big(x_i(k)+q_i-1\big)^+\Big)^2.
\end{aligned}
\end{equation}

\begin{definition}\label{def: MWT-policy}
In each slot $k$, the ARM policy chooses the action that has the maximum weight  in  Table of Fig. \ref{actiontable}.
\begin{figure}[t!]
\centering
\begin{tabular}{|c|c|}
\hline 
$\!\!\!A(k)\!\!\!$& Weights\\
\hline 
$Q_{i,\varnothing}$&$(1-\epsilon_i)\Big(\beta_i\delta_{i,\varnothing}(k)+\lambda f_i(k)\Big)$ \\
\hline
$\oplus_{u\in[l]} Q_{\tau_u,\mathcal{S}_{\tau_u}}$&
$\sum_{u=1}^{l}\beta_{\tau_u} \delta_{\tau_u,\mathcal{S}_{\tau_u}}(k)(1-\epsilon_{\tau_u})$\\
&$+\lambda \sum_{u=1}^l(1-\epsilon_{\tau_u})f_{\tau_u}(k)$ \\
\hline
\end{tabular}
\caption{Coding actions and their weights.}
\label{actiontable}
\end{figure}	
\end{definition}
\begin{remark}\label{remark: buffer size 1}
	When AoI is the only metric in decision making (i.e., $q_i=0$ for all $i$), only the latest packets matters (see Lemma~\ref{lem: at most 1 packet in Qii}). We can thus assume that the buffer size of every queue is $1$ and the stability region is  $\{\theta_i\leq1, i=1,2,\cdots,M\}$.
\end{remark}
\begin{remark}
	We have observed in  simulations that a good approximation of the above ARM policy is obtained by choosing the maximal clique size $l$  be $2$.
\end{remark}
\begin{theorem}\label{thm: MWT-policy-throughput}
	The ARM policy defined in Definition \ref{def: MWT-policy} minimizes the one-slot Lyapunov Drift in each slot.
\end{theorem}
\begin{proof}
The proof of Theorem~\ref{thm: MWT-policy-throughput} is given in Appendix~\ref{App: thm: MWT-policy-throughput}.
\end{proof}

Let  $C^{uncoded}$ be the set of all tuples $\underline{q}$ for which $x_i^+(k)$ is strongly stabilized using the considered coding actions. Upper bounding $J(\underline{q})$ for the ARM policy with a specific choice of $\beta$, we prove the following result in Appendix~\ref{App: thm: upper bound-1}.
\begin{theorem}\label{thm: upper bound-1}	For any $\underline{q}\in C^{uncoded}$, we have the following upper bound on $J(\underline{q})$:
\begin{equation}\label{eq: upper bound2}
J(\underline{q})\leq  \frac{1}{M}\sum_{i=1}^{M}\big(\frac{\alpha_i}{\theta_i}+\frac{\alpha_i}{q_i(1-\epsilon_i)}\big)+\lambda.
\end{equation}
\end{theorem}

\section{Lower Bound}\label{sec: lower bound}

In prior works \cite{cxrAoI12019, IKASEM2019, cxrAoI22019},  lower bounds were found on  AoI as a function of the communication rate.  Similar to \cite[Section~III]{cxrAoI22019}, we derive two lower bounds on the achievable age. The first lower bound is derived by assuming that there is always a fresh packet to be delivered. The second one assumes that all  packets are delivered instantaneously upon  arrivals.

\begin{theorem}\label{thm: lower bound}
For any  policy $\pi$ with communication rate $r_i^\pi$, we have the following lower bounds on $J^\pi(\underline{q})$ in \eqref{eq: tradeoff_objective}: 
\begin{align}
J^\pi(\underline{q})&\geq  \frac{M}{2\sum_{i=1}^{M}r_i^\pi/\alpha_i}+\sum_{i=1}^{M}\frac{\alpha_i}{2M}\label{eq:lb1}\\
J^\pi(\underline{q})&\geq \frac{1}{M}\sum_{i=1}^{M}\frac{\alpha_i}{\theta_i}
\end{align}
\end{theorem}

\begin{proof}
The proof of Theorem~\ref{thm: lower bound} is given in Appendix~\ref{App: thm: lower bound}.
\end{proof}

\begin{corollary}\label{cor: symmetric systems}
For symmetric networks with independent erasure events, the lower bound in \eqref{eq:lb1} leads to
\begin{align*}
J^\pi(\underline{q})\geq \frac{M}{2\epsilon(M)\sum_{i=1}^{M}1/\alpha_i}+\sum_{i=1}^{M}\frac{\alpha_i}{2M}.
\end{align*}
where $\epsilon(M) = \sum_{j=1}^{M}1/(1-\epsilon^j)$.
\end{corollary}
\begin{proof}
The proof of Corollary~\ref{cor: symmetric systems} is given in Appendix~\ref{App: symmetric systems}.
\end{proof}

\section{Numerical Results and Discussion}\label{sec: numerical results}
Finally, we seek to answer the questions that we raised  in Section~\ref{sec: Introduction} through simulations. We assume a symmetric networks with $\epsilon_i=\epsilon$, $q_i = q$, and $\theta_i=\theta$ for $i\in[M]$. 

\subsection{Benefits of Coding}

\begin{figure}[ht!]
\centering
\begin{tikzpicture}[scale=0.7]
\begin{axis}
[axis lines=left,
width=2.9in,
height=2.6in,
scale only axis,
xlabel=$\epsilon$,
ylabel=EAoI,
xmin=0.1, xmax=1,
ymin=3.5, ymax=75,
xtick={},
ytick={},
ymajorgrids=true,
legend style={at={(0.01,0.7)},anchor=west},
grid style=dashed,
scatter/classes={
a={mark=+, draw=black},
b={mark=star, draw=black}
}
]

\addplot[color=blue, dashed]
coordinates{(0.1,11.8023)(0.2,12.5739)(0.3,13.6292)(0.4,15.2723)(0.5,17.8339)(0.6,21.6437)(0.7,27.5448)(0.8,39.0287)(0.9,74.6662)
};

\addplot[color=blue, thick]
coordinates{(0.1,11.7514)(0.2,12.4738)(0.3,13.5129)(0.4,14.9375)(0.5,17.3601)(0.6,20.7712)(0.7,26.4824)(0.8,37.8768)(0.9,72.4053)
};

\addplot[color=brown, dashed]
coordinates{(0.1,8.4519)(0.2,9.4821)(0.3,10.7536)(0.4,12.5636)(0.5,14.8288)(0.6,18.2799)(0.7,24.2866)(0.8,35.9739)(0.9,71.0826)
};

\addplot[color=brown, thick]
coordinates{(0.1,8.4335)(0.2,9.3845)(0.3,10.6832)(0.4,12.2213)(0.5,14.5635)(0.6,18.0863)(0.7,23.5129)(0.8,34.8244)(0.9,67.3896)
};

\addplot[color=green, dashed]
coordinates{(0.1, 7.6726)(0.2, 8.6055)(0.3, 9.8515)(0.4, 11.4836)(0.5, 13.6573)(0.6, 17.2036)(0.7, 22.9309)(0.8, 34.1193)(0.9, 68.5801)
};

\addplot[color=green, thick]
coordinates{(0.1, 7.6791)(0.2, 8.5749)(0.3, 9.7781)(0.4, 11.3536)(0.5, 13.6716)(0.6, 16.9346)(0.7, 22.2920)(0.8, 33.4101)(0.9, 66.3450)
};

\addplot[color=yellow, thick]
coordinates{(0.1, 5.0092)(0.2,5.1560)(0.3, 5.3726)(0.4, 5.6997)(0.5, 6.2108)(0.6, 7.0529)(0.7, 8.5643)(0.8, 11.7519)(0.9, 21.6304)
};

\addplot[color=purple, thick]
coordinates{(0.1, 4)(0.2, 4)(0.3, 4)(0.4, 4)(0.5, 4)(0.6, 4)(0.7, 4)(0.8, 4)(0.9, 4)
};

\legend{Time-sharing when $\theta=0.2$, AMW when $\theta=0.2$,  Time-sharing when $\theta=0.5$, AMW when $\theta=0.5$,    Time-sharing when $\theta=1$, AMW when $\theta=1$, $LB_1$ when $\theta=0.5$,  $LB_2$ when $\theta=0.5$, $UB$ when $\theta=0.5$}

\end{axis}
\end{tikzpicture}
\caption{EAoI as a function of $\epsilon$ and $\theta$ when $M=6$, the upper bound, and the lower bounds.}
\label{fig_1_1}
\end{figure}
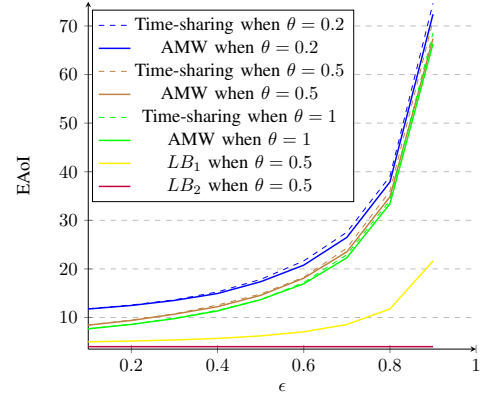

\begin{figure}[ht!]
\centering
\begin{tikzpicture}[scale=0.7]
\begin{axis}
[axis lines=left,
width=2.9in,
height=2.6in,
scale only axis,
xlabel=$\theta$,
ylabel= AoI$_{gap}$,
xmin=0.1, xmax=1,
ymin=0, ymax=1.4,
xtick={},
ytick={},
ymajorgrids=true,
legend style={at={(0.47,.88)},anchor=west},
grid style=dashed,
scatter/classes={
a={mark=+, draw=black},
b={mark=star, draw=black}
}
]

\addplot[color=red, mark=o, smooth, thick]
coordinates{(.1, 1.3898)(.2, 1.1356)(.3, 0.9480)(.4, 0.8323)(.5, 0.7354)(.6, 0.6423)(.7, 0.5249)(.8, 0.4357)(.9, 0.3459)(1, 0.2566)
};

\addplot[color=black, mark=star, smooth, thick]
coordinates{(.1, 0.6154)(.2, 0.5011)(.3, 0.4504)(.4, 0.4077)(.5, 0.3437)(.6, 0.3055)(.7, 0.2382)(.8, 0.1704)(.9, 0.1478)(1, 0.1027)
};

\addplot[color=blue, mark = +, smooth, thick]
coordinates{(.1,0.3575)(.2, 0.2327)(.3, 0.1892)(.4, 0.1629)(.5, 0.1250)(.6, 0.1002)(.7, 0.0748)(.8, 0.0725)(.9, 0.0694)(1, 0.0559)
};

\legend{AoI$_{gap}$ when $M=9$, AoI$_{gap}$ when $M=6$,  AoI$_{gap}$ when $M=3$}

\end{axis}
\end{tikzpicture}
\caption{AoI$_{gap}$ v.s $\theta$  when $\epsilon=0.6$.}
\label{fig_1_2}
\end{figure}
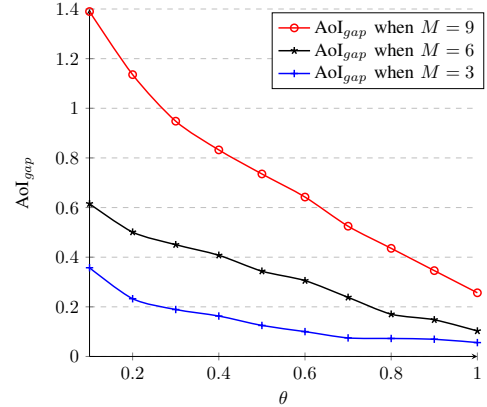

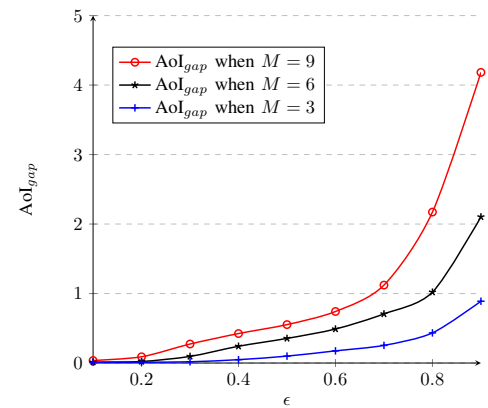
\begin{figure}[ht!]
\centering
\begin{tikzpicture}[scale=0.7]
\begin{axis}
[axis lines=left,
width=2.9in,
height=2.6in,
scale only axis,
xlabel=$\epsilon$,
ylabel= AoI$_{gap}$,
xmin=0.1, xmax=.9,
ymin=0, ymax=5,
xtick={},
ytick={},
ymajorgrids=true,
legend style={at={(0.05,.8)},anchor=west},
grid style=dashed,
scatter/classes={
a={mark=+, draw=black},
b={mark=star, draw=black}
}
]

\addplot[color=red, mark = o, smooth, thick]
coordinates{(.1,0.0366)(.2, 0.0898)(.3, 0.2734)(.4, 0.4242)(.5, 0.5528)(.6, 0.7415)(.7, 1.1214)(.8, 2.1723)(.9, 4.1824)
};

\addplot[color=black, mark = star, smooth, thick]
coordinates{(.1,0.0121)(.2, 0.0218)(.3, 0.0952)(.4, 0.2412)(.5, 0.3540)(.6, 0.4893)(.7, 0.7067)(.8, 1.0197)(.9, 2.1032)
};

\addplot[color=blue, mark = +, smooth, thick]
coordinates{(.1,0.0101)(.2,0.0117)(.3,0.0178)(.4,0.0482)(.5,0.0993)(.6,0.1750)(.7,0.2540)(.8,0.4336)(.9,0.8894)
};

\legend{AoI$_{gap}$ when $M=9$, AoI$_{gap}$ when $M=6$,  AoI$_{gap}$ when $M=3$}

\end{axis}
\end{tikzpicture}
\caption{AoI$_{gap}$ as a function of $\epsilon$  when $\theta=0.2$.}
\label{fig_1_3}
\end{figure}

We first consider the benefits of coding. The ARM policy and the Time-sharing policy are compared in Figure~\ref{fig_1_1} - Figure~\ref{fig_1_3}. To eliminate the impact of rate, we consider the case defined in Remark~\ref{remark: buffer size 1}, i.e., the buffer size of every queue is $1$ and the stability region is $\{\theta\leq1\}$. We have set  $\lambda=0$ and $\beta_i = \min\big\{i, \max\{0,i-3\}\big\}$.
Figure~\ref{fig_1_1} plots the EAoI for $M=6$ users under the ARM and time-sharing policies, and againts the lower bound in \eqref{eq:lb1}. We observe that coding is indeed beneficial when the erasure probability $\epsilon$ is relatively large ($\geq0.6$) and/or the arrival rate $\theta$ is relatively small ($\leq0.5$). When $\theta$ is fixed, EAoI increases with $\epsilon$.

Next, we define AoI$_{gap}$ as the gap between the EAoI under the ARM and time-sharing policies. The relationship between AoI$_{gap}$ and $\theta$ (resp. $\epsilon$) is provided in Figure~\ref{fig_1_2} (resp. Figure~\ref{fig_1_3}).  In Figure~\ref{fig_1_2}, we set $\epsilon=0.6$. We observethat AoI$_{gap}$ (the benfit  of coding) decreases with the arrival rate $\theta$. This is because the (expected) number of newly incoming packets increases with $\theta$ and the availability of fresh uncoded packets weakens the impact of coding actions.

In Figure~\ref{fig_1_3}, we set $\theta=0.2$. AoI$_{gap}$  increases with $\epsilon$. This is because erased packets can be cached and provide more coding opportunities. AoI$_{gap}$ increases slowly when $\epsilon$ is small, and sharply when $\epsilon$ is large.  In addition, from Figure~\ref{fig_1_2} and Figure~\ref{fig_1_3}, the benefits of coding increase with $M$.
\subsection{Impact of Maximal Clique Size}
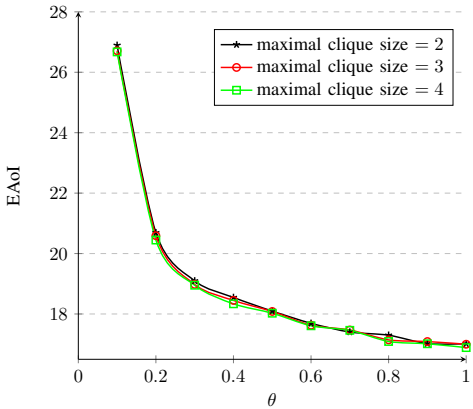
\begin{figure}[ht!]
\centering
\begin{tikzpicture}[scale=0.7]
\begin{axis}
[axis lines=left,
width=2.9in,
height=2.6in,
scale only axis,
xlabel=$\theta$,
ylabel= EAoI,
xmin=0, xmax=1,
ymin=16.5, ymax=28,
xtick={},
ytick={},
ymajorgrids=true,
legend style={at={(0.35,.84)},anchor=west},
grid style=dashed,
scatter/classes={
a={mark=+, draw=black},
b={mark=star, draw=black}
}
]

\addplot[color=black, mark=star, smooth, thick]
coordinates{(0.1,26.8924)(0.2,20.6907)(0.3, 19.0964)(0.4, 18.5474)(0.5, 18.0951)(0.6, 17.6909)(0.7, 17.4026)(0.8, 17.2974)(0.9, 17.0187)(1, 17.01609])
};

\addplot[color=red, mark=o, smooth, thick]
coordinates{(0.1,26.7115)(0.2, 20.5912)(0.3, 18.9790)(0.4, 18.4487)(0.5, 18.0830)(0.6, 17.6266)(0.7, 17.4705)(0.8, 17.1421)(0.9, 17.0878)(1, 16.9987)
};

\addplot[color=green, mark=square, smooth, thick]
coordinates{(0.1,26.6739)(0.2, 20.4469)(0.3, 18.9515)(0.4, 18.3362)(0.5, 18.0256)(0.6, 17.6087)(0.7, 17.4590)(0.8, 17.0893)(0.9, 17.0139)(1, 16.8875])
};

\legend{maximal clique size $=2$,  maximal clique size $=3$, maximal clique size $=4$}

\end{axis}
\end{tikzpicture}
\caption{EAoI under ARM policies under the.}
\label{fig_4_1}
\end{figure}

The impact of maximal clique size is captured in Figure~\ref{fig_4_1}. Let the buffer size of all (virtual) queues be $1$ and set $M=6$, $\lambda=0$, $\epsilon=0.6$ and $\beta_i = \min\big\{i, \max\{0,i-3\}\big\}$. ARM in Definition~\ref{def: MWT-policy} with maximal clique sizes $=2, 3, 4$ are compared in Figure~\ref{fig_4_1}. From Figure~\ref{fig_4_1}, we can see that the ARM policy with maximal clique size of $2$ is a good approximation. This is useful  as it significantly reduces the number of coding actions.

\subsection{Tradeoff between Age and Rate}

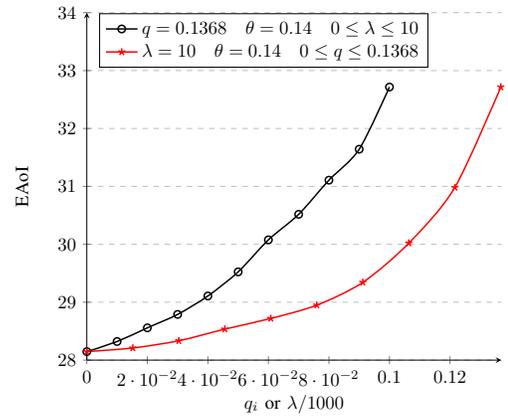
\begin{figure}[ht!]
\centering
\begin{tikzpicture}[scale=0.7]
\begin{axis}
[axis lines=left,
width=3.1in,
height=2.6in,
scale only axis,
xlabel=$q_i$ or $\lambda/1000$,
ylabel= EAoI,
xmin=0, xmax=0.137,
ymin=28, ymax=34,
xtick={},
ytick={},
ymajorgrids=true,
legend style={at={(0.03,0.92)},anchor=west},
grid style=dashed,
scatter/classes={
a={mark=+, draw=black},
b={mark=star, draw=black}
}
]

\addplot[color=black, mark=o, smooth, thick]
coordinates{(0, 28.1465)(0.01, 28.3209)(0.02, 28.5574
)(0.03, 28.7892)(0.04, 29.1076)(0.05, 29.5236)(0.06, 30.0767)(0.07, 30.5174)(0.08, 31.1072)(0.09, 31.6417)(0.1, 32.7156)
};

\addplot[color=red, mark=star, smooth, thick]
coordinates{(0, 28.1465)(0.0152, 28.2093)(0.0304, 28.3342)(0.0456, 28.5350)(0.0608, 28.7191)(0.0760,28.9493)(0.0912,29.3424)(0.1064, 30.0201)(0.1216, 30.9805)(0.1368, 32.7156)
};

\legend{$q=0.1368\quad \theta = 0.14\quad 0\leq\lambda\leq10$, $\lambda=10\quad\theta = 0.14\quad 0\leq q\leq 0.1368$}

\end{axis}
\end{tikzpicture}
\caption{EAoI vs. $q$ and $\lambda$ when $M=3$, $\epsilon=0.6$}
\label{fig_2_1}
\end{figure}

We finally investigate the tradeoff between the AoI and rate. Set $M=3$. The maximum sum-rate achievable with  uncoded caching is around $0.44$, and the channel capacity  is around $0.46$.
Setting $\beta_i = 3$ and $\epsilon=0.6$ in the ARM policy, we first investigate the relationship between $q$ and EAoI (the red star curve in Figure~\ref{fig_2_1}). Now set $\theta=0.14$, $\lambda=10$, $q\in [0, 0.1368]$. EAoI increases with $q$ implying that if the minimum required throughput becomes larger, the system has to sacrifice EAoI to satisfy the rate constraints.   
Next, the relationship between $\lambda$ and EAoI is investigated (the black circle curve in Figure~\ref{fig_2_1}). Let $\theta = 0.14$, $q=0.1368$, $\lambda\in[0,10]$. EAoI increases with $\lambda$. In other words, if the rate constraints become more important, then EAoI increases.  

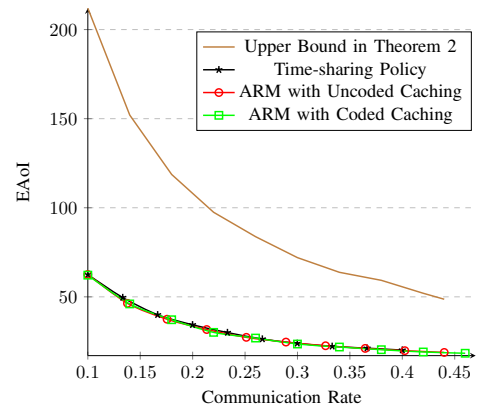
\begin{figure}[ht!]
\centering
\begin{tikzpicture}[scale=0.7]
\begin{axis}
[axis lines=left,
width=2.9in,
height=2.6in,
scale only axis,
xlabel=Communication Rate,
ylabel= EAoI,
xmin=0.1, xmax=0.47,
ymin=17, ymax=212,
xtick={},
ytick={},
ymajorgrids=true,
legend style={at={(0.28,.79)},anchor=west},
grid style=dashed,
scatter/classes={
a={mark=+, draw=black},
b={mark=star, draw=black}
}
]

\addplot[color=brown, mark=smooth, thick]
coordinates{(0.1, 212)(0.14, 152)(0.18, 118.6666)(0.22, 97.4546)(0.26, 83.7792)(0.3, 72)(0.34, 63.7648)(0.38, 59.2632)(0.42, 52)(0.44, 48.6522)
};

\addplot[color=black, mark=star, smooth, thick]
coordinates{(0.1,62.4817)(0.1333, 49.5990)(0.1666, 39.9309)(0.1999, 34.3129)(0.2332, 29.9556)(0.2665, 26.3125)(0.2998, 23.8759)(0.3331, 22.2097)(0.3664, 21.1839)(0.4, 20.1172)
};

\addplot[color=red, mark=o,smooth, thick]
coordinates{(0.1, 62.5396)(0.1378, 46.3498)(0.1756, 37.4344)(0.2134, 31.6520)(0.2512, 27.2650)(0.2890, 24.6124)(0.3268, 22.5492)(0.3646, 21.0586)(0.4024, 19.7896)(0.44, 18.7969)
};

\addplot[color=green, mark=square,smooth,thick]
coordinates{(0.1, 62.1740)(0.14, 46.0229)(0.18, 37.1485)(0.22, 30.0573)(0.26, 26.8651)(0.3, 23.5081)(0.34, 21.8337)(0.38, 20.3396)(0.42, 19.0529)(0.46,18.3391)
};

\legend{Upper Bound in Theorem~\ref{thm: upper bound-1}, Time-sharing Policy,  ARM with Uncoded Caching, ARM with Coded Caching}

\end{axis}
\end{tikzpicture}
\caption{EAoI v.s rate (when $M=3$) under time-sharing, ARM with uncoded and coded caching polices when $\epsilon=0.6$.}
\label{fig_3_1}
\end{figure}

Finally, in Figure~\ref{fig_3_1}, the EAoI is plotted as a function of the communication rate under the
time-sharing policy as well as the ARM policy with uncoded and coded caching. This plot is obtained by setting $\beta_i = \min\big\{i, \max\{0,i-3\}\big\}$, $\theta= q$, and $\lambda = 1$. We observe that  EAoI decreases as rate increases. The three policies have similar performances up to the rate they support. It appears that ARM with coded caching outperforms for rates close to the boundary of the capacity region.

\section{Conclusion}
We investigated the benefit of coding in memoryless broadcast channel with $M$ users. A scheduling framework is proposed in which coding actions, as opposed to users, are scheduled to attain desired tradeoffs between rate and age of information (AoI). 
Two general lower bounds and the upper bound for the proposed MW policies are obtained. Simulation showed that (i) coding is beneficial, and the benefits increase with the number of users; (ii) The tradeoff between rate and AoI exists, EAoI increases with the target rate constraint; (iii) A good approximation of ARM based on maximum clique size was proposed.

\bibliographystyle{IEEEtran}
\bibliography{references}

\appendices

\section{Time-sharing Policies}\label{sec: Benchmark}
We devise deterministic policies without coding using techniques from Lyapunov Optimization. Denote the EAoI for Max-Weight policies (in the long run) as $\mathbb{E}[J]$. Denote $$S(k)=\Big(\{h_i(k)\}_i, \{w_{i,\varnothing}(k)\}_{i}, \{x_i(k)\}_i\Big).$$
Define the Lyapunov function
\begin{align}
L\big(S(k)\big) = \sum_{i=1}^{M} \beta_ih_i(k) + \lambda\sum_{i=1}^{M}\big(x_i^+(k)\big)^2,
\end{align}
where $\beta_i, \lambda>0$,
and the one-slot Lyapunov Drift is defined as
\begin{equation}
\Theta(k) = \mathbb{E}\Big[L\big(S(k+1)\big)-L\big(S(k)\big)|S(k)\Big].
\end{equation}

We  devise the Max-Weight (MW) policy such that it minimizes the one-slot Lyapunov drift:
\begin{definition}\label{def: MW-policy-benchmark}
Let $f_i(k)$ be defined in \eqref{eq: f_i}. In each slot $k$, the MW policy chooses the action that has the maximum weight as shown in following Table:
\begin{figure}[h!]
\centering
\begin{tabular}{|c|c|}
\hline 
$\!\!\!A(k)\!\!\!$& Weights\\
\hline 
$Q_{i,\varnothing}$&$\beta_i\delta_{i,\varnothing}(k)(1-\epsilon_i)+\lambda(1-\epsilon_i)f_i(k)$ \\
\hline
\end{tabular}
\end{figure}	
\end{definition}
\begin{theorem}\label{thm: MW-policy-benchmark}
The MW policy defined in Definition \ref{def: MW-policy-benchmark} minimizes the one-slot Lyapunov Drift in each slot.
\end{theorem}
\begin{proof}
Based on the definition of $t_{i,\varnothing}(k)$, we have
$t_{i,\varnothing}(k)\in\{0,1\}$
and
\begin{align*}
\sum_{i=1}^Mt_{i,\varnothing}(k) = 1.
\end{align*} 
Let $\Theta(k) = \Theta_1(k)  +\lambda\Theta_2(k)$, where
\begin{align*}
\Theta_1(k)=&\mathbb{E}\Big[\sum_{i=1}^{M}\beta_ih_i(k+1) - \sum_{i=1}^{M}\beta_ih_i(k)|S(k)\Big]\\
\Theta_2(k)=&\mathbb{E}\Big[\sum_{i=1}^{M}\big(x_i^+(k+1)\big)^2 - \sum_{i=1}^{M}\big( x_i^+(k)\big)^2 |S(k) \Big].
\end{align*}

We first consider $\Theta_1(k)$.
From  \eqref{eq: recursion of h},
\begin{equation}\label{equ: MW-recursion of h noncoding}
\begin{aligned}
&h_i(k+1)=t_{i,\varnothing}(k)d_i(k)\big(w_{i,\varnothing}(k)+1\big)\\
&\quad+\big(1-  t_{i,\varnothing}(k)d_i(k)\big)\big(h_i(k)+1\big).
\end{aligned}
\end{equation}
Similar with \eqref{eq: queue length}, we have $d_i(k) = H_it_{i,\varnothing}(k)$. 
Using \eqref{equ: MW-recursion of h}, we can re-write the Lyapunov Drift as follows:
\begin{equation*}
\begin{aligned}
\Theta_1(k)=&\sum_{i=1}^{M}\beta_i \mathbb{E}\Big[ H_it_{i,\varnothing}(k)\big(w_{i,\varnothing}(k)-h_i(k)\big)+1|S(k)\Big]\\
=&-\sum_{i=1}^{M}\beta_i \mathbb{E}\big[H_it_{i,\varnothing}(k)\delta_{i,\varnothing}(k)|S(k)\big]+\sum\beta_i\\
=&-\sum_{i=1}^{M}\beta_i(1-\epsilon_i)  \mathbb{E}\big[t_{i,\varnothing}(k)|S(k)\big]\delta_{i,\varnothing}(k)+\sum\beta_i.
\end{aligned}
\end{equation*}
Therefore,
\begin{equation}\label{eq: Theta_age-based noncoding}
\begin{aligned}
\Theta_1(\vec{h}(k))=&\sum_{i=1}^{M}\beta_i-\sum_{i=1}^{M} 1_{\{A(k)=Q_{i,\varnothing}\}}\beta_i\delta_{i,\varnothing}(k)(1-\epsilon_i).
\end{aligned}
\end{equation}

Then, we consider $\Theta_2(k)$. Given $S(k)$, it is sufficient to consider
\begin{align*}	
\tilde{\Theta}_2(k)=\mathbb{E}\Big[\sum_{i=1}^{M}\big(x_i^+(k+1)\big)^2|S(k)\Big].
\end{align*}
Note that 
\begin{equation}
\small
\begin{aligned}
&\mathbb{E}[\big(x_i^+(k+1)\big)^2|S(k)] = \mathbb{E}\Big[\Big(\big(x_i(k)+q_i - d_i(k)\big)^+\Big)^2|S(k)\Big]\\
=&\left\{\begin{aligned}
&\big((x_i(k)+q_i)^+\big)^2&\text{if}&\quad t_{i,\varnothing}(k)=0\\
&\big((x_i(k)+q_i-1)^+\big)^2(1-\epsilon_i)\\
&+\big((x_i(k)+q_i)^+\big)^2\epsilon_i&\text{if}&\quad t_{i,\varnothing}(k)=1
\end{aligned}
\right..
\end{aligned}
\end{equation}

If $A(k) = Q_{i,\varnothing}$, then
\begin{equation*}
\footnotesize
\begin{aligned}
\tilde{\Theta}_2(k) =&(1-\epsilon_i)\bigg(\Big(\big(x_i(k)+q_i-1\big)^+\Big)^2-\Big(\big(x_i(k)+q_i\big)^+\Big)^2\bigg)\\
+& \sum_{j=1}^{M}\Big(\big(x_j(k)+q_j\big)^+\Big)^2. 
\end{aligned}
\end{equation*}

From \eqref{eq: Theta_age-based noncoding} and \eqref{eq: rategain}, minimizing $\Theta(k)$ is equivalent to maximize
\begin{equation*}
\begin{aligned}
&\sum_{i=1}^{M} 1_{\{A(k)=Q_{i,\varnothing}\}}\Big(\beta_i\delta_{i,\varnothing}(k)(1-\epsilon_i)+\lambda(1-\epsilon_i)f_i(k)\Big).
\end{aligned}
\end{equation*}
\end{proof}

\section{Proof of Lemma~\ref{lem: at most 1 packet in Qii}}\label{App: lem: at most 1 packet in Qii}
For any $\mathcal{S}\subset[M]\backslash i$, we first consider $\mathcal{S} =\varnothing$ (Step 1), then consider $\mathcal{S}\neq\varnothing$ (Step 2).

\noindent{\bf Step 1}. Consider $\mathcal{S} =\varnothing$. If $p_j$ is delivered to user $i$ in time slot $k$, then from Definition~\ref{def: AoI for queue}, the AoI of user $i$ is $\min\{k-k_j, h_i(k-1)+1\}$. Note that $k-k_1>k-k_2$, hence 
\begin{equation*}
\small
\begin{aligned}
\min\{k-k_1,h_i(k-1)+1\}\geq\min\{k-k_2,h_i(k-1)+1\},
\end{aligned}
\end{equation*}
which implies $p_1$ provides a larger AoI reduction (for user $i$) than that of $p_2$. Then, transmitting $p_2$ can not be worse than transmitting $p_1$ in terms of AoI. If there is no delivery, then transmitting $p_2$ is again not worse that transmitting $p_1$.

\noindent{\bf Step 2}. Consider $\mathcal{S} \neq\varnothing$. Without loss of generality, suppose that $i, j_1, j_2, \cdots, j_l$ form a maximal clique. Denote the corresponding coded packet as $c_\tau$, $\tau\in\{1,2\}$, respectively. Note that $c_1$ and $c_2$ provide the same AoI reduction to users $ j_1, j_2, \cdots, j_l$. We only consider the case where $c_1,c_2$ can be fully decoded at user $i$; otherwise the AoI of user $i$ does not change after recovering $c_1, c_2$. 

{\bf (1)} If the coded packet is delivered to user $i$, then from Definition~\ref{def: AoI for queue}, the AoI of user $i$ is $\min\{k-k_j, h_i(k-1)+1\}$ when recovering $c_j$. Note that $k-k_1>k-k_2$, so $\min\{k-k_1,h_i(k-1)+1\}\geq\min\{k-k_2,h_i(k-1)+1\}$, $c_2$ provides smaller AoI for user $i$. Then, encoding $p_2$ can not be worse than encoding $p_1$ in terms of AoI.

{\bf (2)} In the future time slot, suppose that $p_2$ has been recovered, and $i, j_1',j_2',\cdots,j_m'$ forms another maximal clique, and the corresponding coded packet is denoted by $c'$.  If $c'$ is delivered to user $i$, by Definition~\ref{def: AoI for queue}, $c'$ can not provide AoI reduction for user $i$. Therefore, the coded packet encoded by uncoded packets from $j_1',j_2',\cdots,j_m'$ provides the same AoI reduction (as $c'$) for users $j_1',j_2',\cdots,j_m'$. 

Thus, from {\bf (1)} and {\bf (2)}, encoding $p_2$ can not be worse than encoding $p_1$.

\section{Proof of Theorem~\ref{thm: MWT-policy-throughput}}\label{App: thm: MWT-policy-throughput}
Based on the definition of $t_{i,\mathcal{S}_i}(k)$, we have
$t_{i,\mathcal{S}_i}(k)\in\{0,1\}$,
and if $A(k)=Q_{i,\mathcal{S}_i}\oplus_{j\in[l]} Q_{\tau_j,\mathcal{S}_{\tau_j}}$, then $t_{i,\mathcal{S}_i}(k)=1$ and $t_{i,\tilde{\mathcal{S}}_i}(k)=0$ where $\tilde{\mathcal{S}}_i\neq\mathcal{S}_i$. Then,
\begin{align}\label{eq: sum t <=1}
\sum_{\mathcal{S}_i\subset[M]\backslash i}t_{i,\mathcal{S}_i}(k)\leq 1.
\end{align} 
Let $\Theta(k) = \Theta_1(k)  +\lambda\Theta_2(k)$, where
\begin{align*}
\Theta_1(k)=&\mathbb{E}\Big[\sum_{i=1}^{M}\beta_ih_i(k+1) - \sum_{i=1}^{M}\beta_ih_i(k)|S(k)\Big]\\
\Theta_2(k)=&\mathbb{E}\Big[\sum_{i=1}^{M}\big(x_i^+(k+1)\big)^2 - \sum_{i=1}^{M}\big( x_i^+(k)\big)^2 |S(k) \Big].
\end{align*}

The relationship between $d_i(k)$ and $t_{i,\mathcal{S}}$ is
\begin{equation}\label{eq: queue length}
\begin{aligned}
d_i^\pi(k) = E_i\sum_{\mathcal{S}\subset[M]\backslash i}t_{i,\mathcal{S}}^\pi(k),
\end{aligned}
\end{equation}
where $E_i$ is a binary random variable independent of $t_{i,\mathcal{S}}^\pi(k)$ for all $\mathcal{S}$ and $k$, $\Pr(E_i=1)=1-\epsilon_i$ and $\Pr(E_i=0)=\epsilon_i$. Thus,
\begin{align}\label{eq: queue length-1}
\mathbb{E}[d_i^\pi(k)] = (1-\epsilon_i)\sum_{\mathcal{S}\subset[M]\backslash i}\mathbb{E}[t_{i,\mathcal{S}}^\pi(k)].
\end{align}

We first consider $\Theta_1(k)$.
Using \eqref{eq: queue length}, \eqref{eq: recursion of h} and \eqref{eq: sum t <=1},
\begin{equation}\label{equ: MW-recursion of h}
\begin{aligned}
&h_i(k+1)=\sum_{\mathcal{S}\subset[M]\backslash i} t_{i,\mathcal{S}}(k)E_i\big(w_{i,\mathcal{S}}(k)+1\big)\\
&\quad+\big(1- \sum_{\mathcal{S}\subset[M]\backslash i} t_{i,\mathcal{S}}(k)E_i\big)\big(h_i(k)+1\big).
\end{aligned}
\end{equation}
Recall that $t_{i,\mathcal{S}_i}(k)\in\{0,1\}$ and $\sum_{\mathcal{S}_i\subset[M]\backslash i}t_{i,\mathcal{S}_i}(k)\leq 1$, so $t_{i,\mathcal{S}_i}^2(k) = t_{i,\mathcal{S}_i}(k)$ and $t_{i,\mathcal{S}_{i}}(k)t_{i,\tilde{\mathcal{S}}_i}(k)=0$ if $\mathcal{S}_i\neq \tilde{\mathcal{S}}_i$. Using \eqref{equ: MW-recursion of h}, we can re-write the Lyapunov Drift as follows:
\begin{equation*}
\small
\begin{aligned}
\Theta_1(k)=&\sum_{i=1}^{M}\beta_i \mathbb{E}\left[\sum_{\mathcal{S}_i\subset[M]\backslash i} E_it_{i,\mathcal{S}_i}(k)\big(w_{i,\mathcal{S}_i}(k)-h_i(k)\big)+1|S(k)\right]\\
=&-\sum_{i=1}^{M}\beta_i \mathbb{E}\big[\sum_{\mathcal{S}_i\subset[M]\backslash i} E_it_{i,\mathcal{S}_i}(k)\delta_{i,\mathcal{S}_i}(k)|S(k)\big]+\sum\beta_i\\
=&-\sum_{i=1}^{M}\beta_i(1-\epsilon_i) \sum_{\mathcal{S}_i\subset[M]\backslash i} \mathbb{E}\big[t_{i,\mathcal{S}_i}(k)|S(k)\big]\delta_{i,\mathcal{S}_i}(k)+\sum\beta_i.
\end{aligned}
\end{equation*}
Therefore,
\begin{equation}\label{eq: Theta_age-based}
\begin{aligned}
\Theta_1(\vec{h}(k))=&\sum_{i=1}^{M}\beta_i-\sum_{i=1}^{M} 1_{\{A(k)=Q_{i,\varnothing}\}}\beta_i\delta_{i,\varnothing}(k)(1-\epsilon_i)\\
-&\sum_{l=2}^{M}\sum_{\tau_1,\tau_2,\cdots,\tau_l}1_{\{A(k)=\oplus_{u\in[l]}Q_{\tau_j,\mathcal{S}_{\tau_u}}\}}\\
\times&\sum_{u=1}^l\beta_{\tau_u}\delta_{i,\mathcal{S}_{\tau_u}}(k)(1-\epsilon_{\tau_u}).
\end{aligned}
\end{equation}

Then, we consider $\Theta_2(k)$. Given $S(k)$, it is sufficient to consider
\begin{align*}	
\tilde{\Theta}_2(k)=\mathbb{E}\Big[\sum_{i=1}^{M}\big(x_i^+(k+1)\big)^2|S(k)\Big].
\end{align*}
Note that 
\begin{equation}
\footnotesize
\begin{aligned}
&\mathbb{E}[\big(x_i^+(k+1)\big)^2|S(k)] = \mathbb{E}\Big[\Big(\big(x_i(k)+q_i - d_i(k)\big)^+\Big)^2|S(k)\Big]\\
=&\left\{\begin{aligned}
&\big((x_i(k)+q_i)^+\big)^2&\text{if}&\sum_{\mathcal{S}_i\subset[M]\backslash i}t_{i,\mathcal{S}_i}(k)=0\\
&\big((x_i(k)+q_i-1)^+\big)^2(1-\epsilon_i)\\
&+\big((x_i(k)+q_i)^+\big)^2\epsilon_i&\text{if}&\sum_{\mathcal{S}_i\subset[M]\backslash i}t_{i,\mathcal{S}_i}(k)=1
\end{aligned}
\right..
\end{aligned}
\end{equation}
Then, we consider the following two cases:

\noindent{\bf Case 1}. If $A(k) = Q_{i,\varnothing}$, then
\begin{equation*}
\footnotesize
\begin{aligned}
\tilde{\Theta}_2(k) =&(1-\epsilon_i)\bigg(\Big(\big(x_i(k)+q_i-1\big)^+\Big)^2-\Big(\big(x_i(k)+q_i\big)^+\Big)^2\bigg)\\
+& \sum_{j=1}^{M}\Big(\big(x_j(k)+q_j\big)^+\Big)^2. 
\end{aligned}
\end{equation*}

\noindent{\bf Case 2}. If $A(k) = \oplus_{j\in[l]}Q_{\tau_j,\mathcal{S}_{\tau_j}}$, then
\begin{equation*}
\footnotesize
\begin{aligned}
\tilde{\Theta}_2(k) =&\sum_{u}^{l}(1-\epsilon_{\tau_u})\bigg(\Big(\big(x_{\tau_u}(k)+q_{\tau_u}-1\big)^+\Big)^2-\Big(\big(x_{\tau_u}(k)+q_{\tau_u}\big)^+\Big)^2\bigg)\\
+& \sum_{j=1}^{M}\Big(\big(x_j(k)+q_j\big)^+\Big)^2. 
\end{aligned}
\end{equation*}
Therefore, 
\begin{equation}\label{eq: rategain}
\begin{aligned}
\tilde{\Theta}_2(k)=&\sum_{j=1}^{M}\Big(\big(x_j(k)+q_j\big)^+\Big)^2-\sum_j^M1_{\{A(k)=Q_{j,\varnothing}\}}f_j(k) \\
-&\sum_{l=2}^{M}\sum_{\tau_1,\tau_2,\cdots,\tau_l}1_{\{A(k)=\oplus_{u\in[l]}Q_{\tau_u,\mathcal{S}_{\tau_u}}\}}\sum_{u=1}^lf_{\tau_u}(k)
\end{aligned}
\end{equation}
From \eqref{eq: Theta_age-based} and \eqref{eq: rategain}, minimizing $\Theta(k)$ is equivalent to maximize
\begin{equation*}
\begin{aligned}
&\sum_{i=1}^{M} 1_{\{A(k)=Q_{i,\varnothing}\}}(1-\epsilon_i)\Big(\beta_i\delta_{i,\varnothing}(k)+\lambda f_i(k)\Big)\\
+&\sum_{l=2}^{M}\sum_{\tau_u, u\in[l]}1_{\{A(k)=\oplus_{u\in[l]}Q_{\tau_u,\mathcal{S}_{\tau_u}}\}}\\
\times&\sum_{u=1}^l(1-\epsilon_{\tau_u})\Big(\beta_{\tau_u}\delta_{i,\mathcal{S}_{\tau_u}}(k)+\lambda f_{\tau_u}(k)\Big).
\end{aligned}
\end{equation*}

\section{Proof of Theorem~\ref{thm: upper bound-1}.}\label{App: thm: upper bound-1}
{\it In the proof, we consider $3$ users. The process can be easily generalized to $M$ users.}

We first consider a stationary randomized policy, 
\begin{align*}
&\Pr(A(k) = Q_{i,\varnothing}) = \mu_{i,\varnothing}\\
&\Pr\{A(k)=\oplus_{j\in[l]}Q_{\tau_j,\mathcal{S}_{\tau_j}}\} = \mu_{(\tau_1, \mathcal{S}_{\tau_1},\cdots,\tau_l,\mathcal{S}_{\tau_l})}.
\end{align*}
Then,
\begin{align}\label{eq: sum=1}
\sum_{j=1}^{M}\mu_{i,\varnothing} + \sum_{l=2}^{M}\sum_{\tau_1, \mathcal{S}_{\tau_1},\cdots,\tau_l,\mathcal{S}_{\tau_l}}\mu_{(\tau_1, \mathcal{S}_{\tau_1},\cdots,\tau_l,\mathcal{S}_{\tau_l})}=1.
\end{align}
As defined before, let $\sigma(\mathcal{I})$ be the probability such that a packet is erased by users in $\mathcal{I}$, and is cached by users in $[M]\backslash\mathcal{I}$. In particular, if the system is independent and symmetric ($\epsilon_1=\cdots = \epsilon_M = \epsilon$), then $\sigma(\mathcal{I}) = (1-\epsilon)^{M-|\mathcal{I}|}\epsilon^{\mathcal{I}}$.

From the proof of Theorem~\ref{thm: MWT-policy-throughput}, 
\begin{equation*}
\begin{aligned}
\Theta_1(k)=&\sum_{i=1}^{M}\beta_i-\sum_{i=1}^{M} \mu_{i,\varnothing}\beta_i\delta_{i,\varnothing}(k)(1-\epsilon_i)\\
-&\sum_{l=2}^{M}\sum_{\tau_1, \mathcal{S}_{\tau_1},\cdots,\tau_l,\mathcal{S}_{\tau_l}}\mu_{(\tau_1, \mathcal{S}_{\tau_1},\cdots,\tau_l,\mathcal{S}_{\tau_l})}\\
\times&\sum_{j=1}^l\beta_{\tau_j}\delta_{i,\mathcal{S}_{\tau_j}}(k)(1-\epsilon_{\tau_j}).
\end{aligned}
\end{equation*}
From definitions of $w_{i,\mathcal{S}}(k)$, we have $w_{i,\mathcal{S}}(k)\leq h_i(k)$ for all $i$ and $\mathcal{S}$. Then, $\mathbb{E}[\delta_{i,\mathcal{S}}(k)]\leq0$. Therefore, remove the terms related to $\mathbb{E}[\delta_{i,\mathcal{S}}(k)]$ when $\mathcal{S}\neq\varnothing$,
\begin{align*}
&\Theta_1(k)\leq \sum_{i=1}^{M}\beta_i-\sum_{i=1}^{M} \mu_{i,\varnothing}\beta_i\delta_{i,\varnothing}(k)(1-\epsilon_i).
\end{align*}
From \cite[Eqn.(60) - (62)]{IKASEM2019}, 
\begin{align*}
\big(x_i^+(k+1)\big)^2 - \big(x_i^+(k)\big)^2\leq& -2x_i^+(k)\big(d_i(k) - q_i\big) + 1\\
\leq& -2x_i^+(k)\big(1 - q_i\big) + 1,
\end{align*}
thus
\begin{equation*}
\small
\begin{aligned}
\Theta_2(k) \leq \sum_{i=1}^M-2x_i^+(k)\big(1 - q_i\big) + M.
\end{aligned}
\end{equation*}
Therefore,
\begin{align*}
&\sum_{k=1}^{K}\mathbb{E}[\Theta(k)]\leq \sum_{k=1}^{K}\sum_{i=1}^M\beta_i(1-\epsilon_i)\mu_{i,\varnothing}\mathbb{E}[w_{i,\varnothing}(k)-h_i(k)]\\
&-\sum_{i=1}^M2\lambda\mathbb{E}[x_i^+(k)](1-q_i) + K\sum_{i=1}^M\beta_i +KM\lambda.
\end{align*}
By algebra, we have
\begin{equation}\label{eq: inequality}
\footnotesize
\begin{aligned}
&\sum_{k=1}^{K}\sum_{i=1}^M\beta_i(1-\epsilon_i)\mu_{i,\varnothing}\mathbb{E}[h_i(k)]+\sum_{k=1}^{K}\sum_{i=1}^M2\lambda\mathbb{E}[x_i^+(k)](1-q_i)\\
&\leq -\sum_{k=1}^{K}\mathbb{E}[\Theta(k)]+\sum_{k=1}^{K}\sum_{i=1}^M\beta_i(1-\epsilon_i)\mu_{i,\varnothing}\mathbb{E}[w_{i,\varnothing}(k)]\\
&+ K\sum_{i=1}^M\beta_i +KM\lambda.
\end{aligned}
\end{equation}
From \eqref{eq: recursion of w_ii}, $w_{i,\varnothing}(k)\leq W_i(k)$, where $W_{i}(k)$ is defined in
\begin{equation}\label{eq: recursion of W}
\begin{aligned}
W_{i}(k+1)=\left\{
\begin{aligned}
&0&\quad& G_i(k)=1\\
&W_{i}(k)+1&\quad&G_i(k)=0
\end{aligned}
\right..
\end{aligned}
\end{equation}
Substituting \eqref{eq: recursion of W} into \eqref{eq: inequality},
\begin{equation}\label{eq: inequality1}
\begin{aligned}
L_1+L_2\leq L_3+\sum_{k=1}^{K}\sum_{i=1}^M\beta_i(1-\epsilon_i)\mu_i\mathbb{E}[W_{i}(k)]+ C.
\end{aligned}
\end{equation}
where 
\begin{align*}
&L_1=\sum_{k=1}^{K}\sum_{i=1}^M\beta_i(1-\epsilon_i)\mu_{i,\varnothing}\mathbb{E}[h_i(k)]\\
&L_2 = \sum_{k=1}^{K}\sum_{i=1}^M2\lambda\mathbb{E}[x_i^+(k)](1-q_i)\\
&L_3 = -\sum_{k=1}^{K}\Theta(k),\quad C = K\sum_{i=1}^M\beta_i +KM\lambda.
\end{align*}

Let $K\to\infty$, 
\begin{align*}
\lim_{K\to\infty}\frac{L_3}{K} \leq \lim_{K\to\infty}\frac{L(S(1))}{K} = 0.
\end{align*}
Thus, dividing by $KM$ on both sides of \eqref{eq: inequality1}, we have
\begin{equation}\label{eq: inequality2}
\footnotesize
\begin{aligned}
&\lim_{K\to\infty}\frac{L_1}{K}+\lim_{K\to\infty}\frac{L_2}{K}\\
&\leq \frac{\sum_{i=1}^M\beta_i(1-\epsilon_i)\mu_{i,\varnothing}\sum_{k=1}^K\mathbb{E}[W_{i}(k)]}{KM} + \frac{\sum_{i=1}^M\beta_i}{M}+\lambda.
\end{aligned}
\end{equation}
Note that $W_i(k)$ is a geometric random variable with parameter $\theta_i$, we have
\begin{equation}\label{eq: mid upper bound3}
\begin{aligned}
&\lim_{K\to\infty}\frac{L_1}{K}+\lim_{K\to\infty}\frac{L_2}{K}\\
\leq&\frac{1}{M}\sum_{i=1}^{M}\beta_i\big(\frac{(1-\epsilon_i)\mu_{i,\varnothing}}{\theta_i}+1\big)+\lambda.
\end{aligned}
\end{equation}
Since $L_1>0$, so \eqref{eq: mid upper bound3} is reduced to
\begin{equation}\label{eq: mid upper bound4}
\begin{aligned}
&\sum_{i=1}^{M}2\lambda(1-q_i)\lim_{K\to\infty}\frac{\sum_{k=1}^{K}\mathbb{E}[x_i^+(k)]}{K}\\
&\leq \frac{1}{M}\sum_{i=1}^{M}\beta_i\big(\frac{(1-\epsilon_i)\mu_{i,\varnothing}}{\theta_i}+1\big)+\lambda.
\end{aligned}
\end{equation}
Then,
\begin{align*}
\lim_{K\to\infty}\frac{\sum_{k=1}^{K}\mathbb{E}[x_i^+(k)]}{K}<\infty,\quad i=1,2,\cdots,M.
\end{align*}
In fact, strong stability of the process $x_i^+(k)$, i.e.,
\begin{align*}
\lim_{K\to\infty}\frac{1}{K}\sum_{k=1}^{K}\mathbb{E}\big[x_i^+(k)\big]<\infty,
\end{align*}
is sufficient to establish that the minimum rate constraint, $r_i^\pi\geq q_i$, is satisfied \cite{IKASEM2019}, \cite[Theorem~2.8]{M2010}.

Since $L_2>0$, so
\begin{align}\label{eq: final}
\lim_{K\to\infty}\frac{L_1}{MK}\leq \frac{1}{M}\sum_{i=1}^{M}\beta_i\big(\frac{(1-\epsilon_i)\mu_{i,\varnothing}}{\theta_i}+1\big)+\lambda.
\end{align}

\begin{figure}[t!]
	\centering
	\includegraphics[width=2in, height = 1.8in]{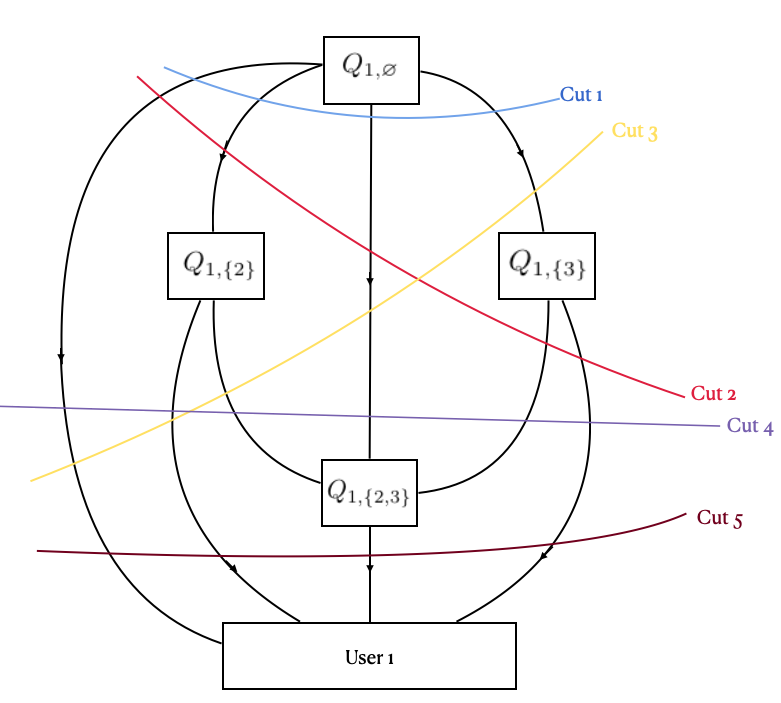}
	\caption{The flow of user $1$ in the virtual network.}
	\label{fig:cut}
\end{figure}

Now, we consider the probabilities of actions, $\{\mu_{i,\varnothing}\}$ and $\{\mu_{\tau_1,\mathcal{S}_{\tau_1},\cdots,\tau_l,\mathcal{S}_{\tau_l}}\}$. Consider $M=3$. Note that $(q_1, q_2, q_3)$ gives the minimum requirements of rate, then for each $i$. In the graph of virtual network, every user has $5$ cuts, see Figure~\ref{fig:cut}. Take user~$1$ as an example,
Cut~$1$ implies that
\begin{equation}\label{eq: flow network1}
\footnotesize
\begin{aligned}
\mu_{1,\varnothing}(1-\epsilon_1) + \sum_{j\in\{2,3\}}\mu_{1,\varnothing}\sigma([3]\backslash j) + \mu_{1,\varnothing}\sigma(\{1\})\geq q_1,
\end{aligned}
\end{equation}
Cut~$2$ implies that
\begin{equation}\label{eq: flow network2}
\footnotesize
\begin{aligned}
&\mu_{1,\varnothing}(1-\epsilon_1) + \mu_{1,\varnothing}\sigma(\{1,3\}) + \mu_{1,\varnothing}\sigma(\{1\}) \\
&+ \big(\mu_{1,\{3\},3,\{1\}} + \mu_{1,\{3\},3,\{1,2\}}\big)\big(1-\epsilon_1 + \sigma(\{1\})\big) \geq q_1,
\end{aligned}
\end{equation}
Cut~$3$ implies that
\begin{equation}\label{eq: flow network3}
\footnotesize
\begin{aligned}
&\mu_{1,\varnothing}(1-\epsilon_1) + \mu_{1,\varnothing}\sigma(\{1,2\}) + \mu_{1,\varnothing}\sigma(\{1\}) \\
&+ \big(\mu_{1,\{2\},2,\{1\}} + \mu_{1,\{2\},2,\{1,3\}}\big)\big(1-\epsilon_1 + \sigma(\{1\})\big) \geq q_1,
\end{aligned}
\end{equation}
Cut~$4$ implies that
\begin{equation}\label{eq: flow network4}
\footnotesize
\begin{aligned}
&\mu_{1,\varnothing}(1-\epsilon_1) +  \mu_{1,\varnothing}\sigma(\{1\}) \\
&+ \big(\mu_{1,\{3\},3,\{1\}} + \mu_{1,\{3\},3,\{1,2\}}\big)\big(1-\epsilon_1 + \sigma(\{1\})\big)\\
&+ \big(\mu_{1,\{2\},2,\{1\}} + \mu_{1,\{2\},2,\{1,3\}}\big)\big(1-\epsilon_1 + \sigma(\{1\})\big) \geq q_1,
\end{aligned}
\end{equation}
Cut~$5$ implies that
\begin{equation}\label{eq: flow network}
\footnotesize
\begin{aligned}
&\mu_{1,\varnothing}(1-\epsilon_1) +  \big(\mu_{1,\{2,3\}, 2, \{1\}} + \mu_{1,\{2,3\}, 3, \{1\}} \\
&+ \mu_{1,\{2,3\}, 2, \{1,3\}} + \mu_{1,\{2,3\}, 3, \{1,2\}}\big)(1-\epsilon_1) \\
&+ \big(\mu_{1,\{3\},3,\{1\}} + \mu_{1,\{3\},3,\{1,2\}}\big)(1-\epsilon_1)\\
&+ \big(\mu_{1,\{2\},2,\{1\}} + \mu_{1,\{2\},2,\{1,3\}}\big)(1-\epsilon_1) \geq q_1.
\end{aligned}
\end{equation}
Similar process for users $2$ and $3$, we get \eqref{eq: cut1} $\sim$ \eqref{eq: cut5},
\begin{equation}\label{eq: cut1}
\footnotesize
\begin{aligned}
&\mu_{i,\varnothing}(1-\epsilon_i) + \sum_{j\in[3]\backslash i}\mu_{i,\varnothing}\sigma([3]\backslash j) + \mu_{i,\varnothing}\sigma(\{i\})\geq q_i,
\end{aligned}
\end{equation}
\begin{equation}\label{eq: cut2}
\footnotesize
\begin{aligned}
&\mu_{i,\varnothing}(1-\epsilon_i) + \mu_{i,\varnothing}\sigma([3]\backslash j) + \mu_{i,\varnothing}\sigma(\{i\}) \\
&+ \big(\mu_{i,\{k\},k,\{i\}} + \mu_{i,\{k\},k,[3]\backslash k}\big)\big(1-\epsilon_i + \sigma(\{i\})\big) \geq q_i\\
&j, k\neq i, \,\, j,k\in[3],\,\, j\neq k,
\end{aligned}
\end{equation}
\begin{equation}\label{eq: cut4}
\footnotesize
\begin{aligned}
&\mu_{i,\varnothing}(1-\epsilon_i) +  \mu_{i,\varnothing}\sigma(\{i\}) \\
&+\sum_{j\in [3]\backslash i} \big(\mu_{i,\{j\},j,\{i\}} + \mu_{i,\{j\},j,[3]\backslash j}\big)\big(1-\epsilon_i + \sigma(\{i\})\big) \geq q_i,
\end{aligned}
\end{equation}
\begin{equation}\label{eq: cut5}
\footnotesize
\begin{aligned}
&\mu_{i,\varnothing}(1-\epsilon_i) +  \sum_{j\in [3]\backslash i}\big(\mu_{i,[3]\backslash i, j, \{i\}} +  \mu_{i,[3]\backslash i, j, [3]\backslash j}\big)(1-\epsilon_i) \\
&+\sum_{j\in [3]\backslash i} \big(\mu_{i,\{j\},j,\{i\}} + \mu_{i,\{j\},j,[3]\backslash j}\big)(1-\epsilon_i) \geq q_i.
\end{aligned}
\end{equation}

Thus, $\{\mu_{i,\varnothing}\}$ and $\{\mu_{\tau_1,\mathcal{S}_{\tau_1},\cdots,\tau_l,\mathcal{S}_{\tau_l}}\}$ satisfying \eqref{eq: sum=1}, \eqref{eq: cut1} $\sim$ \eqref{eq: cut5} give the inequality \eqref{eq: final}. Note that $\sum_{i=1}^{3} q_i<1$. So let $\mu_{i,\varnothing}=q_i$, we can choose proper $\{\mu_{\tau_1,\mathcal{S}_{\tau_1},\cdots,\tau_l,\mathcal{S}_{\tau_l}}\}$ such that  \eqref{eq: sum=1}, \eqref{eq: cut1} $\sim$ \eqref{eq: cut5} are satisfied. Substituting $\mu_{i,\varnothing} = q_i$ into \eqref{eq: final}, we get 
\begin{align*}
&\frac{1}{3}\sum_{i=1}^{3}(1-\epsilon_i)q_i\cdot\lim_{K\to\infty}\frac{1}{K}\sum_{k=1}^{K}\beta_i\mathbb{E}[h_i(k)]\\
\leq&\frac{1}{3}\sum_{i=1}^{3}\beta_i\big(\frac{(1-\epsilon_i)q_i}{\theta_i}+1\big)+\lambda.
\end{align*}
In particular, set $\beta_i = \frac{\alpha_i}{(1-\epsilon_i)q_i}$, which yields to
\begin{align*}
J(\underline{q})\leq\frac{1}{3}\sum_{i=1}^{3}\big(\frac{\alpha_i}{\theta_i}+\frac{\alpha_i}{q_i(1-\epsilon_i)}\big)+\lambda.
\end{align*}

\section{Proof of Theorem~\ref{thm: lower bound}}\label{App: thm: lower bound}
For a large time horizon $K$ and look at the packets intended for user $i$. Let $N_i(K)$ denote the number of  recovered packets up to and including time slot $K$. Now consider the $m^{th}$ and $(m+1)^{th}$ recovered packets and denote the delivery time of them at user $i$ by $T_i(m)$ and $T_i(m+1)$, respectively. The inter-delivery time
\begin{align}\label{eq: inter delivery}
I_i(m) = T_i(m+1) - T_i(m)
\end{align} 
is the time between these two consecutive deliveries. Upon arrival of the $m^{th}$ recovered packet for user $i$, the age of information of user $i$ drops to the value $D_i(m)$ which represents how much delay the packet has experienced in the system. Let $L_i$ be the number of remaining time slots after the last packet recovery for user $i$. Now define $\Gamma_i(m)$ as the sum of age functions $h_i(k)$, where $k$ is in the interval $[T_i(m), T_i(m+1))$:
\begin{align}
\Gamma_i =& \sum_{k=T_i(m)}^{T_i(m)+I_i(m)-1}h_i(k)\\
=&\frac{1}{2}I_i^2(m)-\frac{1}{2}I_i(m)+D_i(m-1)I_i(m)\label{eq: Gamma}.
\end{align}
It follows that in the limit of large $K$, we have
\begin{align}\label{eq: AoI Gamma}
J^\pi(M) = \lim_{K\to\infty}\Big[\frac{1}{M}\sum_{i=1}^{M}\alpha_i\frac{1}{K}\sum_{m=1}^{N_i(K)}\Gamma_i(m)\Big].
\end{align}
Using this formulation, we next lower bound EAoI. Let $R_i$ denote the capacity outer bound  of user $i$, defined in \eqref{eq: outer bound1} and \eqref{eq: outer bound2}. Note that in the limit of large $K$, $\frac{N_i(K)}{K}$ is the rate/throughput of user $i$ under policy $\pi$, denoted by $R_i^\pi$.

Then, we prove the first part of Theorem~\ref{thm: lower bound}.
Consider any scheduling policy and a large time-horizon $K$. The EAoI can be re-written in terms of $\Gamma_i(m)$:
\begin{equation*}
\footnotesize
\begin{aligned}
J^\pi_K = \frac{1}{M}\sum_{i=1}^{M}\frac{\alpha_i}{K}\big(\sum_{m=1}^{N_i(K)} \Gamma_i(m)+\frac{1}{2}L_i^2+D_i(N_i(K))L_i-\frac{1}{2}L_i\big).
\end{aligned}
\end{equation*}
Since $D_i(m)\geq 1$ for all $1\leq m\leq N_i(K)$, we can lower bound \eqref{eq: AoI Gamma} by substituting $D_i(m-1)=1$. Using similar steps as \cite[Eqns. (12) - (17)]{IKASEM2018},
\begin{align}\label{eq: lowerbound1}
\lim_{K\to\infty}\mathbb{E}\big[J^\pi_K]\geq\lim_{K\to\infty}\mathbb{E}\big[\frac{1}{2M}\sum_{i=1}^{M}\frac{\alpha_iK}{N_i(K)}+\frac{\sum_{i=1}^{M}\alpha_i}{2M}\big].
\end{align}
Now note that by the Cauchy-Schwarz inequality, we have
\begin{align*}
\lim_{K\to\infty}\mathbb{E}\Big[\sum_{i=1}^{M}\frac{N_i(K)}{\alpha_iK}\Big]\mathbb{E}\Big[\sum_{i=1}^{M}\frac{\alpha_iK}{N_i(K)}\Big]\geq M^2,
\end{align*}
and thus
\begin{align}
\lim_{K\to\infty}\mathbb{E}\big[\sum_{i=1}^{M}\frac{\alpha_iK}{N_i(K)}\big]\geq M^2/\sum_{i=1}^{M}\frac{R_i}{\alpha_i}.
\end{align}
Inserting this back into \eqref{eq: lowerbound1}, we obtain
\begin{align*}
J^\pi(M)\geq\frac{M}{2\sum_{i=1}^{M}\frac{R_i}{\alpha_i}}+\frac{\sum_{i=1}^{M}\alpha_i}{2M}.
\end{align*}

Next, we prove the second part of Theorem~\ref{thm: lower bound}. Suppose that all packets are recovered instantaneously with one time-unit delay. A lower bound to EAoI in this scenario constitutes a lower bound to EAoI in our setup. Let $X_i(m)$ denote the inter arrival time between $m^{th}$ and $(m+1)^{th}$ packets. $\big\{
X_i(m)\big\}_m$ is a geometric i.i.d sequence. Under the assumption of instantaneous recovery, $I_i(m)=X_i(m)$. It hence follows from \eqref{eq: Gamma} that 
\begin{align*}
\Gamma_i(m)=\frac{1}{2}X_i^2(m)+\frac{1}{2}X_i(m).
\end{align*}
Thus, similar with \cite{cxrAoI22019}, the time-average AoI of user $i$ is
\begin{align*}
\mathbb{E}[h_i] =& \lim_{K\to\infty}\frac{1}{K}\sum_{k=1}^{K}h_i(k)\\
=&\frac{\mathbb{E}[X_i^2(m)]}{2\mathbb{E}[X_i(m)]}+\frac{1}{2}=\frac{2-\theta_i}{2\theta_i}+\frac{1}{2}
\end{align*}
and
\begin{align*}
J^\pi(M)\geq \frac{1}{M}\sum_{i=1}^{M}\frac{\alpha_i}{\theta_i}.
\end{align*}

\section{Proof of Corollary~\ref{cor: symmetric systems}}\label{App: symmetric systems}
The proof consists of minimizing \eqref{eq:lb1} over all rate tuples $(r^\pi_1,\ldots,r^\pi_M)\in\hat{\mathcal{C}}$ where $\hat{\mathcal{C}}$ is an outer bound on the capacity region \cite[Section~III]{MGLT2013}.

From \cite[Section~III]{MGLT2013}, denote $\hat{C}$ as the channel capacity outer bound. Let $\pi$ is a permutation of $[M]$ such that $\pi(M-i+1)=\hat{\pi}(i)$, where $\hat{\pi}$ is the permutation defined in \cite[Definition~1]{MGLT2013}. Recall that $\sigma(\mathcal{I})$ is  the probability that an erasure occurs for all users in $\mathcal{I}$. Denote $\hat{\epsilon}_{\pi(i)}=\sigma(\cup_{j=1}^{i}\{\pi(j)\})$. From \cite[Lemma~3, Lemma~4]{MGLT2013}, we can obtain the outer bound $\hat{C}$ 
\begin{align}\label{eq: outer bound1}
\hat{C} = \cap_\pi \hat{C}_{\pi}
\end{align}
where
\begin{align}\label{eq: outer bound2}
\hat{C}_{\pi} = \Big\{(R_1,R_2,\cdots,R_M)|\sum_{i\in[M]}\frac{R_{\pi(i)}}{1-\hat{\epsilon}_{\pi(i)}}\leq1 \Big\}.
\end{align}
Consider symmetric and independent channels, i.e., $\epsilon_i=\epsilon$ for all $i\in[M]$. From \eqref{eq: outer bound2}, the capacity outer bound $\hat{\mathcal{C}}$ is given by
\begin{align}\label{eq: capacity}
\hat{\mathcal{C}}=\Big\{(R_1,\cdots,R_M)|0\leq R_i\leq \frac{1}{\sum_{j=1}^{M}1/(1-\epsilon^j)}\Big\}.
\end{align}
From \eqref{eq: capacity}, $LB_1^\pi$ can be re-written as
\begin{align*}
LB_1^\pi = \frac{M}{2\epsilon(M)\sum_{i=1}^{M}1/\alpha_i}+\sum_{i=1}^{M}\frac{\alpha_i}{2M}.
\end{align*}
where $\epsilon(M) = \frac{1}{\sum_{j=1}^{M}1/(1-\epsilon^j)}$.

\end{document}